# What Is Actually Being Annotated?

# Inter-Prompt Reliability as a Measurement Problem in LLM-Based Social Science Labeling

Jingyuan Liu

Boston University

**Abstract**

Large language models (LLMs) are increasingly used for annotation in computational social science, yet their methodological reliability under prompt variation remains unclear. This paper introduces Inter-Prompt Reliability (IPR), a framework for evaluating the stability of LLM outputs across semantically equivalent but linguistically varied prompts. Drawing on Inter-Rater Reliability, IPR is measured by Pairwise Agreement Rate (PAR) and its distribution to capture both consistency and stochasticity in model behavior. We evaluate this framework on two tasks with distinct properties: TREC (interpretative) and Politifact (knowledge-anchored). Results show that LLM annotation exhibits substantial stochastic variation in interpretative tasks, while appearing more stable in knowledge-based tasks. We further show that majority voting across prompts significantly improves reproducibility and reduces variance. These findings suggest that LLM prompt acts as an instrumental measurement while its wording exhibits methodological uncertainty. For future LLM-based CSS studies, we suggest that researchers moves beyond single-prompt evaluation toward distributional stability and prompt aggregation within our IPR framework.

# Introduction

Large language models (LLMs) are increasingly being used in Computational Social Science (CSS) research to analyze and annotate text data (Gilardi, Alizadeh and Kubli 2023; Liu and Sun 2023). As the accessibility and efficacy of LLM improve with monthly updating models, the shift is particularly outstanding in computational social science, where annotations are used to conduct qualitative research on abstract social concepts into measurable variables. (Deterding and Waters 2018)

Unlike in natural language processing (NLP), where labels are often treated as end outputs of predictive optimization, in social science, annotations serve as an instrumental proxy to identify a construct or concept within the text (Grimmer & Stewart, 2013; Kleinheksel et al., 2020). As emphasized by Krippendorff (2018), the scientific value of a label lies not in its existence, but in its ability to facilitate stable social inference. Consequently, reliability, especially Inter-Rater Reliability (IRR) (McHugh, 2012) is not just a technical metric but a prerequisite for valid measurement. This leads to a fundamental, yet unaddressed tension in LLM-based annotation: inter-prompt reliability (IPR).

We define our IPR as the stability of model outputs across semantically and format equivalent prompts that vary only in wording. In traditional manual annotation, IRR focuses more on human consensus. When diffentiation across annotators exist, it is often treated as inherent noise. However, LLM-based annotation introduces a systematically new source of variance, prompt formulation. Since the same semantic prompt can be written in infinite linguistic variations by different prompt-writer, annotation outcomes naturally form a distribution across prompts. Accordingly, IPR should be evaluated not just by its mean performance, but also as a stochastic distribution characterized by its Standard Deviation (SD). Hypothetically, if annotation shifts when a prompt is reworded by a different researcher without changing its sematic content, a critical ontological question arises: What is actually being annotated? The conceptual construct that we intend to discover, or linguistic phrasing of the prompt itself?

Prior work in NLP has identified prompt as a key variant, being greatly influenced by wording, casing, spacing, formatting(Melanie et al., 2024) and example-strategy. However, it is treated more as a model performance issue, leading to development of new evaluation metrics like sensitivity and consistency (Federico et al., 2025). However, CSS studies of LLM-based annotations remains relying heavily on singe-prompt, simple-run (Gilardi et al., 2023; Mellon et al., 2024; Hoes et al., 2023; Castro-Gonzalez et al., 2024) design. Such practice may unintentionally lead to a 'prompt cherry-picking' result, adding prompt wording as a new uncontrolled variable. This ignorance of prompt itself as a measurement instrument could lead to serious distorted view of LLM capabilities and compromise replicability of findings.

Thus, our focus is thus not just accuracy of LLM annotation, but whether they perform consistently with different expressions of prompts with sematic information. To address this gap, we ask: (1) To what extent does prompt wording introduce variability in LLM-based annotations and how does this affect their IPR as measurement instruments. (2) Is IPR task dependent? (3) How can such instability be mitigated?

We test two different LLMs, including one of OpenAI's ChatGPT, GPT-4o mini and one open-sourced Llama, LLaMa3.1:8b with 20 semantically same prompts with categorized communication style across two tasks. On a question classification task relying more on interpretability, the TREC dataset, with accuracy ranging from 0.61 to 0.90 and high IPR variance, which can lead to serious questions in research replicability and misleading conclusions in later analysis. On another ground-truth based misinformation identification task, the PolitiFact dataset, the IPR shows a much lower deviation across prompts. This differentiation indicates the prompt sensitivity is strongly task dependency: On interpretative-heavy tasks (Soft) requiring more interpretability and self-evaluation, LLM annotation shows great instability and replicability, while on knowledge-anchored tasks (Hard) task with solid outside knowledge remain relatively robust.

This finding shows the dual nature of LLM-based annotation: the high efficacy and cost-saving of this method come at the cost of high uncontrollability and instability. Consider the nature of qualitative methods, an interpretative analytical approach relying on careful, nuanced coding (Deterding and Waters 2018), such CSS work inevitably falls into the category of interpretative-heavy soft tasks. In this context, relying on LLMs to annotate and code text with single prompt is methodologically dangerous. It is therefore crucial to come up with certain techniques to reduce the inherent stochasticity of LLM outputs, ensuring that automated tools do not compromise the rigor required for sociological inquiry.

To mitigate this issue, we propose a simple yet robust way, majority voting, which aggregates outputs from multiple prompts. This method significantly improves both stability and accuracy in soft tasks. Standard deviation of IPR has been reduced by 90% and mean of accuracy increases approximately 9% when subset of 10 prompts is compared to single-prompt settings. Though increasing 10 times of running time and cost, it still offers a practical trade-off between efficiency and reliability in LLM-based annotation. Despite a higher cost, the overall efficacy of LLM-annotation with majority voting still remains competitive to current methods of manual coding. However, researchers still need to consider the underlying stochasticity of LLM and the forthcoming validation issue in the growing application of computational methods in CSS.

To ensure the clarity across a wide range of backgrounds, including both sociologists and data scientists, Table1 provide a glossary of terms before proceeding to the theoretical background that might be useful to readers.

Table 1. Glossary of Key Terms

| Term | Definition |
| --- | --- |
| | *Qualitative* |
| Annotation | Coded text - text that has a label or code applied to it. |
| Code | Labels applied to annotated text that describe its content. |
| Coding | The process of assigning codes to annotated text. |
| Crowd workers | Using (usually untrained and anonymous) humans to perform tasks, often through online platforms such as MTurk. |
| IRR (Inter-Rater Reliability) | A statistical measure assessing the degree of agreement among multiple human annotators. |
| RTA (Reflexive Thematic Analysis) | A qualitative methodology that treats interpretive variation across analysts as a source of analytical richness rather than error. |
| Instrumental Measurement | A methodological perspective that frames annotation as mediated by an instrument, where variation in outputs may arise from the properties of the instrument itself, challenging the assumption that interpretive differences reflect meaningful analytical insight. |
| | *Language Modelling* |
| Language Modeling | The task of predicting or generating text based on learned probabilistic patterns in language data. |
| Prompt | A textual instruction or input used to elicit a response from an LLM. |
| Prompt Engineering | The practice of designing and refining prompts to optimize model outputs. |
| IPR (Inter-Prompt Reliability) | A measure of consistency of LLM outputs across semantically equivalent but linguistically varied prompts. |
| PAR (Pairwise Agreement Rate) | The proportion of agreement between outputs generated from different prompts, computed pairwise. |
| Stochasticity | The inherent randomness in model outputs due to probabilistic sampling mechanisms. |
| Linguistic Noise | Variability in model outputs introduced by differences in prompt phrasing rather than underlying semantic content. |

## Literature review

Computational methods have been increasingly used to analyze textual data and conduct scientific analysis of behavioral data (Errica et al., 2025) since the emergence of CSS (Lazer et al. 2020). However, annotation has long been a major obstacle that has prevented them from being used more widely (Rao, 2023). Researchers have to conduct original annotations to ensure that the labels match their categories (Benoit et al., 2016). Mostly, these works have been done with expert coders or crowd workers on platforms such as Amazon Mechanical Turk (MTurk), both of which, has to deal with a dilemma within costs, efficacy and quality (Gilardi et al., 2023).

However, this landscape has been reshaped with the emergence of LLMs, especially GPT models. LLMs have been suggested as a potentially promising way to annotate large bodies of text, with some studies speculating that they may transform the social science research (Ziems et al., 2024). Recent papers demonstrate that, for relatively simple coding tasks, LLM-based annotation could offer accurate outcome along with low cost and time efficacy, including automated content moderation (Gilardi et al., 2023), political issues categorization (Mellon et al., 2024), misinformation identification (Hoes et al., 2023), sentiment analysis (Castro-Gonzalez et al., 2024), etc.

More often, literature in social science focuses on LLM performance on a given prompt performs superior to manual annotator in cost and speed. However, this practice may unintentionally rely on 'prompt cherry-picking', leading to unreplicable results. As most studies report performance under a single selected prompt, they leave unanswered whether these results are stable, reproducible, and methodologically trustworthy.

This concern is especially important since LLM outputs are inherently stochasticity. McCoy et al. (2023) showed that LLM performances depend on the likelihood of the input prompt and of the correct output answer. Studies in NLP have long been discussing prompting protocols to restrain or, manage to control the stochastic nature, leading to the study of prompt engineering(White et al., 2023). Many prompt design techniques have been proved to have an ignorable influence on model performance, including role prompting (Kong et al., 2024), chain of thoughts(Wei et al., 2022; Kojima et al., 2022), example-strategy like zero-shot, one-shot, few-shot(Wei et al., 2022; Kojima et al., 2022) and sequences of examples(Zhao et al., 2021).Even prompt formatting, including casing, space, separator could lead to great performance spread of accuracy from 0.036 to 0.804. (Sclar et al., 2024) Besides, model parameters including temperature also have a huge influence(Holtzman et al., 2020), while this does not fall within the scope of this paper because we treat prompt itself as the single variant here.

In the field of NLP, the stochasticity is normally discussed as the robustness of models themselves, being a part of evaluation metrics of model performance as an engineering problem.

While in social science, this problem remains to exert a much deeper influence, since it would directly affect the annotation reliability, replicability of experiments, even the result itself.

Our paper speaks to this literature in questioning whether LLM annotation is truly trustworthy by introducing Inter-Prompt Reliability (IPR) as a measure of stability of LLM-based annotation across semantically equivalent prompts. Inspired by Inter-Rater Reliability (IRR), a golden standard of in CSS, IPR evaluates the internal consistency of LLM-based annotation under semantically same but linguistic differently prompt by measuring pairwise agreement. This variability through the distribution is further characterized by standard deviation.

While Julian et al. (2025) recently explored 'inter-prompt agreement,' their work primarily focused on definition-level information variation, comparing different definition strategies (e.g., no definition, researcher-generated and LLM-generated). Consequently, their metrics serve as an evaluation of different prompt strategy, whereas our IPR isolates sematic invariance, quantifying the stochastic noise produced by linguistic phrasing alone.

By shifting the focus from finding the best prompt towards assessment of LLM-annotation as a scientific instrument, this study establishes a new framework for measurement reliability specifically for LLM-annotation. Thus, we investigate whether LLM-based annotation remains stable with semantic invariance using IPR and its standard deviation. Our finding shows that the stochastic nature of LLM produces significant inter-prompt variation, reduced replicability and reliability in interpretative task. To mitigate this methodological risk, we propose and validate a simple but effective strategy, Majority Rating, to reduce the stochastic noise while also enhances accuracy, thereby stabilizing LLMs to rigorous social science inquiry.

## Method

The goal of our experiment is to assess how sensitive and reliable prompts are in the context of English-based text labeling datasets in the field of social science. Our intent is to analyze the impact of prompt variations from wording and prompting strategies from a completely different angle compared to previous works: we do not talk about accuracy, we talk about inter-prompt reliability.

*Prompt Corpus*

To systematically examine the effect of linguistic variation on prompts, we designed a specialized prompt to maintain semantic invariance while introducing controlled linguistic variation. This further ensures wording remains to be the only variant of any observed fluctuation in model outputs rather than shifts in task information.

The corpus consists of two distinguish style, reflecting diverse ways how researchers interact with LLMs. For each dataset, the corpus contains a total of 20 prompts (2N, where N=10 per style) per dataset, ensuring a statistical distribution of IPR:

Formal/Analytical: Professional, concise and instruction-heavy.

Conversational/Contextual: Natural, narrative and unrefined.

To ensure sematic consistency, all prompts are derived from a standard prompt mentioned in previous literature or accompanied dataset instructions (Hovy et al., 2001; Li and Roth, 2002; Misra, et al., 2022). To stimulate casual rephrasing from different researchers, we applied a surface-level linguistic features, for example, verbosity, sequences of expression, sentence structure. Meanwhile, overall formatting and labelling criteria are preserved. This design allows us to manage to restrain the variant to linguistic variance as much as possible. Example of standard analytical and contextual prompt can be found in Appendix.

*IPR and PAR Framework*

To better interpret the instability of LLM-based annotation, we introduce IPR as a measurement of cross-prompt reliability through Pairwise Agreement Rate (PAR). While IPR functions as the conceptual center to guide the overall process, PAR measures the agreement rate of annotation generated with two different prompts, regardless of the ground truth label.

PAR measures the probability that two prompts assign the same label to the same input. For two distinct prompts $p_i$ and $p_j$ with discrete labels, we define this agreement as:

$$PAR_{i,j} = \frac{1}{N} \sum_{x=1}^{N} \mathbf{1}\left(p_i(x) = p_j(x)\right),$$

where $N$ is the total number of data to annotate, $p_i(x)$ denotes the label assigned by prompt $p_i$ to sample $x$ and $\mathbf{1}(\bullet)$ is the indicator function that equals 1 if the condition is met and 0 otherwise.

For graded labels, the agreement further measures the distance across each category in a more refined way after converted prediction labels into scores $s_i$ and $s_j$:

$$PAR_{i,j} = \frac{1}{N} \sum_{x=1}^{N} 1 - \frac{|s_i(x) - s_j(x)|}{D},$$

where D refers to the max distance across the grades.

This formulation measures the probability that two prompts produce identical outputs over the same dataset.

Given a set of prompts $P$, we propose two ways to capture IPR through the lens of $PAR$, the average agreement across all prompt pairs and standard deviation of $PAR$:

1. Mean PAR: Reflect the average level of agreement of consistency across outputs from different prompts.

$$\mu_{PAR} = \frac{1}{\binom{P}{2}} \sum_{i<j} PAR_{i,j}$$

2. Standard Deviation of PAR: Measures the dispersion of agreement across prompts, capturing the extent of prompt wording-induced stochasticity.

$$\sigma_{PAR} = \sqrt{\frac{1}{\binom{P}{2} - 1} \sum_{i<j} (PAR_{i,j} - \mu_{PAR})^2}$$

This method allows us to distinguish between two aspects of reliability: the mean agreement which reflects overall consistency, and the dispersion of agreement (SD), which captures the extent of prompt-induced stochasticity.

Tradition metrics like accuracy are aggregated metrics that fails to consider the predictive stochasticity inherent in individual interactions. They often focus on the overall end effect, whether the model is 'right' rather than stable. However, in social science research, a measurement instrument that reach the correct answer by chance or through specific linguistic luck is methodologically unsound. By calculating the replicability of LLM outputs across semantically same prompts, PAR offers a more rigorous analysis of whether the model has truly captured the underlying sociological construct or is merely reacting to linguistic noise.

*Dataset*

We consider two distinct multi-class classification datasets for comparison purpose: TREC (Hovy et al., 2001; Li and Roth, 2002) and Politifact (Misra, et al., 2022).

The TREC dataset requires the model to categorize an open-ended question into six discrete class (e.g. Abbreviation, Entity, Human). Since this task relies primarily on syntactic and semantic analysis with relatively low dependence on external knowledge, we define it as a 'Soft Task'. Furthermore, the categories are mutually exclusive and discrete, offering a clean environment to measuring linguistic-induced variance without label or ranking structure ambiguity. The evaluation is conducted on 500 tests samples.

Differently, the Politifact dataset asks the LLM to determine the level of authenticity of a certain tweet. Given that fact-checking remains one of the most prominent ways to identify inauthentic and misleading content (Allen et al., 2021; Walter et al., 2020) in CSS, evaluating reliability of LLMs in this domain is of great importance. This task requires an intrinsic leverage of LLM's internal parametric knowledge, making it inherently knowledge-anchored. Thus, we define it as a 'Hard Task'. We sampled 200 test cases from each of the six ground-truth classes (e.g. True,

False, Half-True, Pants-On-Fire), 1200 samples in total. Notably, these labels are ordinal, allowing us to examine how IPR interacts with graded classification boundaries.

Together, these two datasets create a complementary analysis of IPR under various levels of task complexity. This enables us to discover the boundary condition of LLM stability in real-world social science annotation settings.

Also, since the goal of our experiment is to examine the instability in prompt variance with semantical information, considering previous studies on prompt sensitivity on formatting, structure, and model sensitivity on temperature parameter. We specifically set the temperature = 0 and restrain all the casing, spacing techniques to be exactly the same.

## Results

In this section, we mainly want to answer three questions: (1) is prompt variation (word) a significant variant in LLM output stability even under sematic invariance? (2) whether task type moderates prompt-induced stochasticity and (3) how the linguistic noise of LLM-based annotation can be reduced.

### RQ1: Linguistic Variation on Measurement Stability

To answer question (1), we tested our corpus of 20 semantically equivalent prompts on the TREC datasets to classify the desired answer type. As a "Soft Task" primarily dependent on semantic parsing, TREC result serves as a baseline to separate pure linguistic effects from knowledge-based errors.

Table 2 shows the overall summary of accuracy compared with the grounded truth and variance inside each category along with std. Detailed accuracy score could be found in Appendix. The first observation is that accuracy varies significantly across semantically equivalent prompts, ranging from 0.808 (prompt: Examine the meaning of the question and identify the type of answer the question is expecting from Number, Location, Person, Description, Entity, or Abbreviation.) to 0.546 (prompt: If someone asks the following question, what kind of answer are they expecting from Number, Location, Person, Description, Entity, or Abbreviation?) for GPT-4o mini, and from 0.756 to 0.392 for LLaMa3.1:8b. This large performance spread in accuracy while no-seemingly difference in prompts indicates that linguistic variation alone can induce significant differences in model outputs.

| Model | Prompt Type | Mean Acc | Std Acc | Min | Max |
|---|---|---|---|---|---|
| GPT-4o mini | Analytical | 0.737 | 0.055 | 0.628 | 0.808 |
| | Contextual | 0.699 | 0.076 | 0.546 | 0.796 |
| | Overall | 0.718 | 0.068 | 0.546 | 0.808 |
| LLaMa3.1:8b | Analytical | 0.595 | 0.084 | 0.392 | 0.716 |
| | Contextual | 0.56 | 0.108 | 0.412 | 0.756 |

|  | Overall | 0.578 | 0.097 | 0.392 | 0.756 |

Table 2: Summary Statistics of Prompt-Level Accuracy (TREC)

To further illustrate the internal variability across prompt, Figure 1 presents a PAR heatmap, offering a direct overview how each prompt is aligning with another one. As shown in Figure1, both models show substantial disagreement across semantically equivalent prompts, with PAR spread over 40% This level of variation exceeds commonly accepted thresholds in human annotation settings, where disagreement above approximately 20% is typically considered problematic. While GPT-4o mini (left) shows higher overall consensus (darker blue regions), it still shows instability clots activated by specific prompts (e.g., analytical_6 and contextual_2). To further quantify this variability, we also calculated the standard deviation of PAR, which remains high over 0.1, indicating this deviation is broadly distributed across different prompts, indicating further instability in output generated by semantically same but different wording prompts.

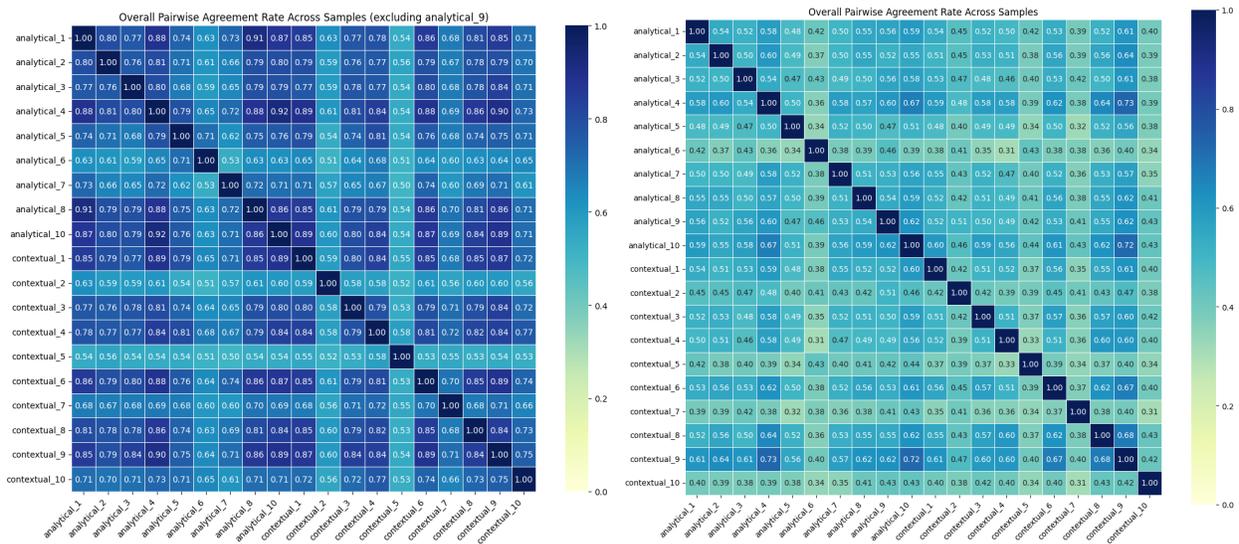

Figure 1: PAR Heatmap of TREC (Left: GPT-4o mini, Right: LLaMa3.1:8b)

Importantly, certain low-accuracy prompts (e.g., analytical_6) showed high PAR (up to 71%) with other prompts. This further indicates that PAR (or IPR) and accuracy capture distinct aspects of model behavior. While PAR reflects consistency across prompts, indicating stability of prompt performance, accuracy reflects correctness with respect to ground truth. As a result, high accuracy does not necessarily imply true ability. It could still be inferenced or mislead by linguistic bias across prompts.

*RQ2: Task Complexity*

We then answer the second question by conducting the same type of experiment with another 20 prompts on a completely different task problem, the Politifact dataset. The Politifact dataset require LLM to identify the extent of accuracy of a certain text and output among six ordinal classes. Different from the TREC task, which we defined as 'soft task' that rely heavily on interpretation, Politifacts belongs to a high anchored 'hard task' focusing more on fact grounded checking, which we believe the judgement are more based on external factual knowledge. We hypothesize that because this task is grounded in the model's internal parametric knowledge, it should exhibit different sensitivity to prompt variation.

To reflect the ordinal nature of PolitiFact labels, we first mapped the ordinal class to a continuous score to preserve the distance between labels. Since labels are converted into score, the so-called accuracy, measured as the normalized distance to ground truth, Closeness, shows a extremely low std of 0.004-0.005 across both models shown in Table 3. Consistently, the PAR heatmap in Figure 2 also reveals a much narrow fluctuation range of only 3% with SD of 0.02, indicating a much high inter-prompt agreement under continuous scoring in Politifact datasets.

| Model | Prompt Type | Mean Closeness | Std Closeness | Min | Max |
|---|---|---|---|---|---|
| GPT-4o mini | Analytical | 0.763 | 0.004 | 0.755 | 0.768 |
| | Contextual | 0.764 | 0.004 | 0.760 | 0.770 |
| | Overall | 0.763 | 0.004 | 0.755 | 0.770 |
| LLaMa3.1:8b | Analytical | 0.712 | 0.005 | 0.706 | 0.721 |
| | Contextual | 0.705 | 0.005 | 0.692 | 0.721 |
| | Overall | 0.709 | 0.006 | 0.692 | 0.721 |

Table 3: Summary Statistics of Prompt-Level Accuracy (Politifact)

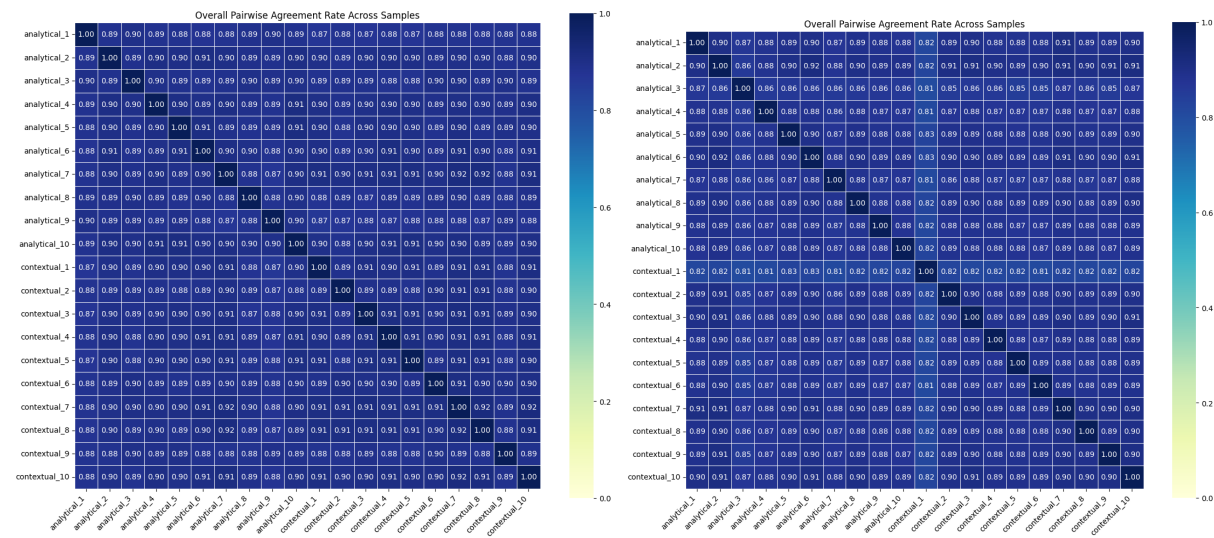

Figure 2: PAR Heatmap of Politifact (Ordinal) (Left: GPT-4o mini, Right: LLaMa3.1:8b)

To eliminate the possible influence of label representation (continuous/discrete), we then further binarize the labels, measuring agreement by identical match (1/0). As expected, the overall accuracy decreases to approximately 30% due to stricter evaluation criteria. While the absolute PAR values are low (ranging from 0.46 to 0.71), the Standard Deviation of PAR still shows a relatively low performance 0.051 and 0.063. This indicates the model has entered a 'stubborn' state, remaining to be wrong while stable. While the model struggles with the factual complexity of the task which leads to lower PAR and accuracy, its responses are prompt invariant.

In the TREC task, the high SD suggested a Stochastic Error that the model was confused by the wording. In contrast, low SD in PolitiFact, both ordinal and discrete labels suggest a systematic bias. Since the task is knowledge-anchored, the LLM relies on its internal parametric memory. This internal anchor acts as a stabilizer that suppresses linguistic noise, leading to a state of 'Stubborn Consistency' that the model reaches the same wrong answer regardless of whether the prompt is formal, conversational, or structurally varied.

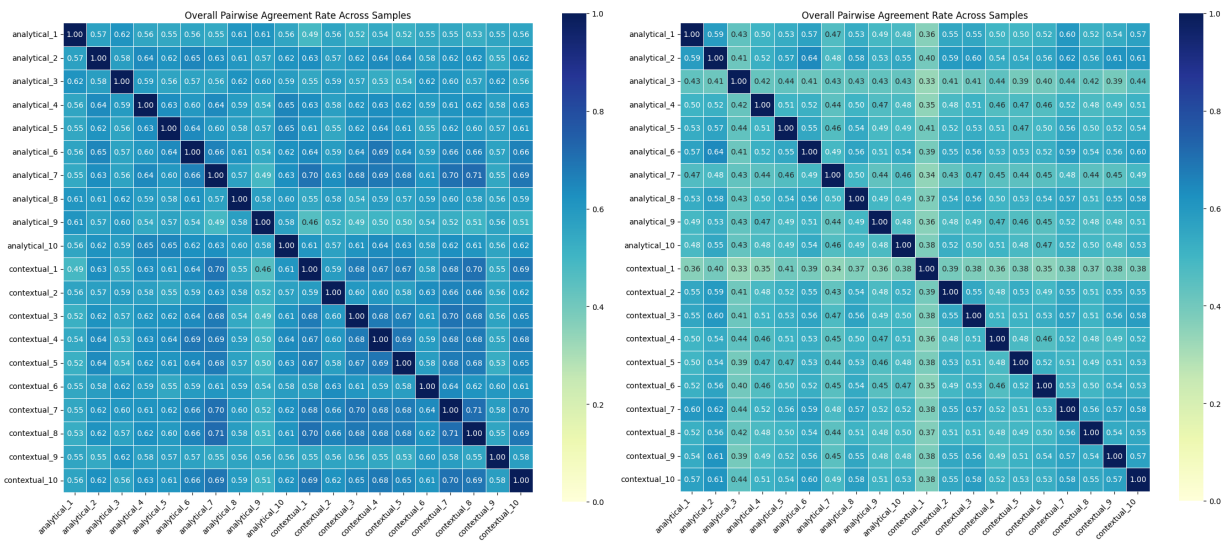

Figure 3: PAR Heatmap of Politifact (Discrete) (Left: GPT-4o mini, Right: LLaMa3.1:8b)

The great deviation of PAR std from these two types of datasets shows a critical finding: the prompt stability on LLM shows a Knowledge Anchoring Effect. Also, task type is a major moderator of prompt sensitivity. When a task is grounded in model's internal factual knowledge and training data, it resists strong stylistic and wording interference. Otherwise, when a task relies solely on the question interpretation, different wording could introduce a relatively high portion of noise, leading to instability in output.

### RQ3: Mitigating through Majority Voting

Given the significant sensitivity of LLM-annotation in interpretative tasks, we answer the third question by introducing Majority Voting as a reliability buffer. Majority Voting functions as an all-to-one approach where the final annotation is determined by $k$ outputs generated from $k$ semantically equivalent but linguistically varied prompts.

Figure 3 shows a great improvement of PAR both in mean and std with increasing number of prompts aggregated into the voting process ($k$=1, 3, 5, 10) in the TREC tasks. As shown, for GPT-4o mini, the mean PAR rise from 0.71 to nearly 0.9 while standard deviation dropped by 0.8 from k=1 to k=5. Beyond overall improvement, we also discover that the variance reduction is particularly significant at a lower value of k, where SD dropped sharply from k=1 to k=3. This indicates even a small ensemble of prompt results could mitigate the stochastic linguistic noise greatly. When k reaches 10, the Mean PAR converges to a highly reliable 0.95, and the SD effectively collapses, further demonstrating the potential of Majority Voting in noise reducing, though accompanied with a saturation effect.

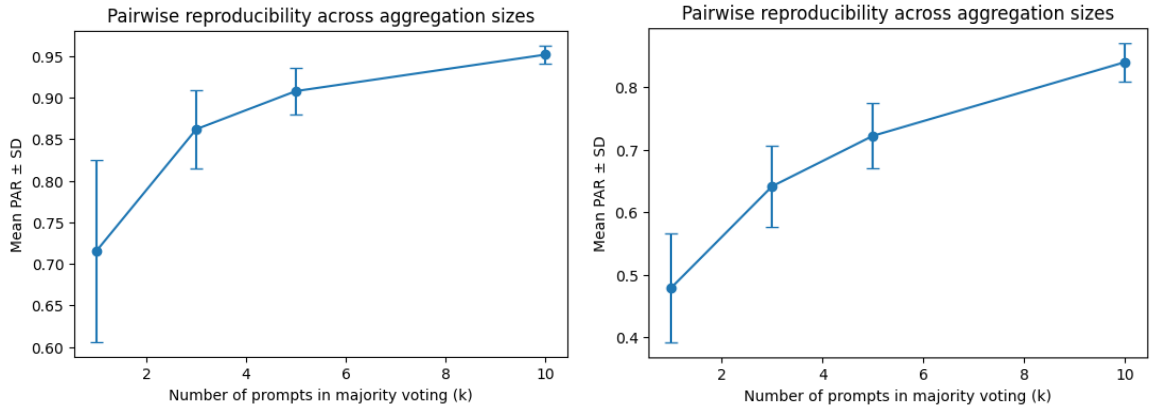

Figure :PAR across aggregation size (Left: GPT-4o mini, Right: LLaMa3.1 8b)

From a methodological perspective, this result has two important implications. First, it demonstrates that the instability observed in single-prompt LLM annotation is not purely random but can be systematically reduced through controlled aggregation. Second, it reframes the possible way of prompt design, that in CSS, rather than searching for a single optimal prompt, researchers may instead rely on prompt ensembles as a more robust annotation strategy.

Nonetheless, though improving consistency, it is important to note that majority voting does not necessarily improve correctness. If the model exhibits systematic bias, as mentioned in previous works by Julian et al. (2025), aggregation may further reinforce consistent but incorrect predictions, leading to what we called stubborn consistency in previous section.

# Discussion

*Prompt as a Measurement Instrument*

The central finding of our study is that LLM based annotation remains highly sensitive to prompt wording, even along with semantically equivalent prompts. This pattern is specifically discovered in soft and interpretive task where the labels depend less on factual knowledge, but more on linguistic framing, category boundary interpretation and latent understanding.

Considering existing literature rely heavily on a single prompt strategy, this finding has important implications in CSS. In human-annotation, disagreement is mostly attributed to differences of theoretical interpretation or coder background. However, in LLM-based annotation, a pure black box, even when the so-called 'annotator' remains to be the same model, the prompt refers to the same meaning, differences in wording can still produce significant deviation in outputs. As a result, researchers who wants to annotate the same sample may obtain different labeling outcome simply because they have different phrasing and wording habits. This instability of annotation may then lead to further distorted patterns, ideology, tendency and final findings.

Thus, we argue for a paradigm shift in CSS, that a prompt should not be viewed merely as a conversational interface, but as a formal measurement instrument. Like a physical scale must be calibrated, LLM prompt sets also need to be tested for instrumental reliability. Furthermore, LLM-based annotation should not be viewed as fixed point estimates, but as distributions across a linguistic prompt space.

*IPR as a CSS Diagnostic Framework*

To conceptualize this problem, we introduced Inner-Prompt Reliability (IPR) as an analogue to IRR. Alike IRR, the goal of IPR remains to test the stability of the overall annotation system. While IRR evaluates the consistency of judgments across human annotators, IPR evaluates the consistency of LLM outputs across semantically equivalent prompts that vary only in wording or style. Both are at an 'annotator' level in the overall system.

To measure IPR, we further introduced Pairwise Agreement Rate (PAR), which measure the agreement between the output generated by two prompts across the same dataset. A standard deviation level of IPR is also computed to capture the deviation of agreement within the prompt sets, offering a reviewing of whether the distribution of agreement is consistent. This serves as a complementary technique to examine whether agreement is consistent or unevenly dependent on specific wording. Additionally, two forms of PAR: discrete PAR (for categorical labels) and continuous PAR (for ordinal scales), whose formula could be found in the Method section. These two indicators make IPR adaptable to different annotation settings.

Together, our finding suggests that prompt variation in wording leads to huge methodological risk that can jeopardize replicability of annotation and further findings. Even when researcher believe that they are asking the same question, there still exists a great possibility that the output remains differently. IPR is therefore proposed not simply as new matric, but as a framework to diagnose whether LLM-based annotation is robust enough to support scientific claims in CSS.

*Risk of LLM-application in Qualitative Research*

Beyond a diagnostic framework of LLM-annotation, our finding has broader implications for growing and heated use of LLMs in qualitative research annotation, particularly in interview-based analysis. These tasks are inherently more interpreted than our soft task, often relying heavily on ambiguous boundaries and context dependent judgment along with long text. If in relatively constrained tasks such as TREC, prompt induced variance already existed in a substantial way, it is reasonable to expect that such variability will be amplified in a more qualitative workflow.

This also introduce an additional layer of methodological risk beyond commonly discussed concerns such as model bias, under-representative training and hallucination. To mitigate this risk, we argue that existed qualitative analysis should incorporate human-in-loop reliability controls, like human verification or auditing. Moreover, rather than relying on a single prompt, researchers should adopt multiple prompt strategies across semantically equivalent formulations, where our simple aggregation-based technique, Majority Voting, could inherently reduce the linguistic noise, while maintain the majority output in a single run. Nonetheless, it still needs to be noticed that majority voting could also led to a stubbornly biased output when the model itself is intrinsically problematic. Therefore, IPR should be used as a measure of stability, not as a substitute for validity.

*Risk of Spurious Novelty in LLM-assisted RTA*

A further implication of our findings shows concerns on newly rising discussion of LLMs in Reflexive Thematic Analysis (RTA). Unlike traditional reliability framework like IRR, RTA embrace the interpretive variations as an analytical product. In RTA framework, disagreement is not necessarily treated as error, but a source of richness and alternative perspective.

Recent studies have begun to extend this logic to LLM-based qualitative workflows. For example, prior work has explored the use of LLMs to support reflexive coding practices and thematic analysis (Vikan, et al., 2026; Dunivin, Z.O., 2025). In these approaches, variation across outputs is often interpreted as complementary of gaps in the researchers' perspectives rather than instability.

However, our findings suggest that this extension needs to be further examined. Human-centered interpretive divergence in RTA framework is assumed to be grounded in different theoretical understanding and disciplinary backgrounds. While in LLM-based RTA, variations may arise not only from difference in framework interpretation, but also from prompt framing itself. Linguistic noise acts as an additional resource of variant that is not reflecting any meaningful theoretical insight. This creates a serious methodological ambiguity that when an LLM produces a novel interpretation, it becomes difficult to determine whether that response reflects a meaningful extension of the theoretical lens or merely an artifact of prompt-level stochasticity.

For this reason, we argued that further LLM-based RTA researchers should incorporate stronger grounding mechanism than existing literature. One possible strategy is to add Chain of Thought (CoT) as a part of prompt engineering to encourage the model explicitly explains how it works. This also provide a more transparent reasoning path, so that researchers can better distinguish grounded interpretation from to stochastic noise. LLMs indeed generate useful interpretive variations distinguished from human understanding, but they may also generate spurious novelty that only appears meaningful due to linguistic behaviors. Therefore, it creates a further challenge that researchers have to further determine which form of diversity is generated by LLMs.

## Limitation

This study presents several limitations that needs further investigation.

First, the prompt set used in this study was written by a single researcher, which limits the diversity of linguistic styles represented in the experiment. Although the prompts were designed to be semantically equivalent, they still reflect one individual's prompting habits, phrasing preferences. In real research settings, however, prompts are often written by different researchers, each bringing different vocabularies, writing habits. As a result, the current design may underestimate the degree of variability that would emerge in real-world valid multi-researcher settings. Future work should therefore expand the prompt sets by involving multiple researchers from different disciplinary backgrounds in CSS to construct semantically equivalent prompts. Such a design would better capture realistic variation in prompt formulation and provide a stronger test of inter-prompt reliability in actual scientific practice.

Second, our focus was limited to relatively structured annotation tasks. Although IPR is already observable under these conditions, it may become substantially more in real-world CSS settings. Social science often involves open-ended data, such as interview transcripts, ethnographic notes, or discourse materials. These forms of data typically involve greater ambiguity and higher interpretive flexibility, all of which may amplify prompt-induced variation. It remains unclear what extent the instability identified through IPR would extend to such settings. Future research should therefore test the IPR framework on interview-based qualitative materials in order to evaluate its robustness under more interpretive research conditions.

Third, we isolated wording while ignoring other variables. Factors like model temperature, decoding strategies, and Chain-of-Thought prompting may also interact with prompt wording. Future research could investigate how these parameters combined affect overall annotation reliability.

# Conclusion

Our study shows the LLM-based annotation does not remain stable and reliable even under semantically equivalent prompts. Minor variation in wording and phrasing could still lead to significant variations in model output, especially in interpretative tasks. This challenges the common assumption that LLM outputs could be replicated by using the same model and task specification.

To address this issue, we introduce Inter-Prompt Reliability as a conceptual framework to evaluate the stability of LLM-based annotation across prompts. IPR is further measured through Pairwise Agreement Rate (PAR) and its distribution. Rather than focusing on identifying a single optimal prompt, our approach reframe LLM prompts as a crucial instrument measurement that exerts great reliability problem.

Broadly, our findings suggest that LLM annotations should not be treated as a deterministic labeling process, but as a distributional and potentially unstable measurement instrument like human annotators. Though strategies like Majority Voting can partially stabilize the final label, they do not eliminate the intrinsic stochasticity induced by prompt variations. As an unstable instrumental measurement, prompt design output, variability, and reliability assessment should be incorporated into the methodology pipeline. Future work may explore strategies to provide solid validations of LLM output and distinguish meaningful interpretive answers from stochastic noise.

Ultimately, as LLMs become more deeply integrated into both quantitative and qualitative workflows, the primary challenge for researchers is to differentiate between meaningful interpretive divergence and stochastic linguistic noise. By adopting our IPR framework, scholars can more rigorously evaluate the stability of their findings, making them replicable evidence rather than a lucky roll of the linguistic choice.

# Appendix

|  | TREC | Politifact |
|---|---|---|
| Standard | Classify the questions based on the answer type from Number, Location, Person, Description, Entity, or Abbreviation. | Can you fact-check a statement for me? When fact-checking, avoid negations and only use one of the following labels to classify each statement:• TRUE – The statement is accurate and there's nothing significant missing.• MOSTLY TRUE – The statement is accurate but needs clarification or additional information.• HALF TRUE – The statement is partially accurate but leaves out important details or takes things out of context.• MOSTLY FALSE – The statement contains an element of truth but ignores critical facts that would give a different impression.• FALSE – The statement is not accurate.• PANTS ON FIRE – The statement is not accurate and makes a ridiculous claim.• NO VERDICT – The statement lacks sufficient context, or there is not enough information to assess the veracity of the statement. |
| Analytical | Identify the kind of information the following question is requesting from Number, Location, Person, Description, Entity, or Abbreviation. | Classify the factual accuracy of the statement using one of the labels. Return only the label:\n\nTRUE – The statement is correct and does not omit any important information.\nMOSTLY TRUE – The statement is generally correct but requires some clarification or additional context.\nHALF TRUE – The statement contains some accurate elements but leaves out key details or misrepresents context.\nMOSTLY FALSE – The statement has a small amount of truth but ignores essential facts that change the whole interpretation.\nFALSE – The statement is incorrect.\nPANTS ON FIRE – The statement is completely false and highly misleading.\nNO VERDICT – There is insufficient information to determine whether the statement is accurate. |
| Contextual | Read the question and determine the kind of answer that best fits its intent from Number, Location, Person, Description, Entity, or Abbreviation. | Can you help me check if this statement is true? Based on your understanding, choose the best label. Return only the label:\n\nTRUE – The statement is correct and nothing important is missing.\nMOSTLY TRUE – The statement is mostly correct but could use more explanation.\nHALF TRUE – The statement has some truth but leaves out key details.\nMOSTLY FALSE – The statement contains a bit of truth but ignores important facts.\nFALSE – The statement is not correct.\nPANTS ON FIRE – The statement is completely false and very misleading.\nNO VERDICT – There isn't enough information to evaluate it. |